\documentclass[11pt,twoside]{article}
\usepackage{asp2014}

\aspSuppressVolSlug
\resetcounters

\begin{document}

\title{Debian Astro: An open computing platform for astronomy}

\author{Ole~Streicher
  \affil{Leibniz Institute for Astrophysics, Potsdam, Germany;
    \email{olebole@debian.org}}}

\paperauthor{Ole~Streicher}{ole@aip.de}{0000-0001-7751-1843}
            {Leibniz Institute for Astrophysics}
            {}{Potsdam}{Brandenburg}{14489}{Germany}

\begin{abstract}
  Debian Astro is a Debian Pure Blend that aims to distribute the available
  astronomy software within the Debian operating system. Using Debian as the
  foundation has unique advantages for end-users and developers such as an easy
  installation and upgrading of packages, an open distribution and development
  model, or the reproducibility due to the standardized build system.
\end{abstract}

\section{Introduction}
In the last years, the amount and quality of software for astronomy
raised significantly. Many traditional software packages got a
significant push.  With Astropy \citep{2013A&A...558A..33A}, a new and
coordinated approach of software development in astronomy was
established.  However, there is still much legacy software in use,
partly based on outdated dependencies and sometimes even not
maintained anymore.

Keeping up a consistent integration of astronomical software into a
single usable framework is difficult, as it is to manage the packages,
their requirements and to provide sensible defaults to the
users. Debian Astro integrates tested software packages, maintained by
a group of volunteers. Along with the packaging, we also provide a web
site with detailed description and status of all
packages\footnote{\url{https://blends.debian.org/astro}}.

The Debian Astro Pure Blend is completely integrated into Debian, so
that the packages are directly installable on any Debian installation
from the regular software repositories.

\section{Packaging for Debian}

Debian as a free software distribution has a number of strict rules
for the inclusion of packages. The Social Contract
\citep{DebianSocialContract} contains the Debian Free Software
Guidelines with licensing requirements, limiting the software to Open
Source. In exceptional cases, non-free software or software with
non-free dependencies may be included in the ``contrib'' and
``non-free'' areas, which are however not an official part of Debian.

All technical requirements for Debian packages are described in the
Debian Policy \citep{DebianPolicy}. There are rules on the package
names, shared libraries, dependency declaration and so on. The rules
are derived from the common practice in software development and
ensure a consistent packaging.

Debian is not only available for Linux at x86 processors, but for a
variety of kernels and CPU types. Aside from the instruction set, the
architectures differ by word width (32 or 64 bit), byte order and
other properties.

Although many Debian packages are maintained by a single person, it is
encouraged to maintain the packages within a team. Teams exist for
packaging software of specific languages, but also by topic, like the
Debian Science team. In 2014, the Debian Astro team was founded,
dedicated to maintain packages that are relevant to astronomy and
astrophysics. The team is organized by a mailing
list\footnote{\url{https://lists.debian.org/debian-astro}}. To join
the team it is not required to be a Debian developer, anyone can
contribute here. Uploads of non-developers must be sponsored by a
Debian developer from the team.

The packaging work is organized in git repositories located on the
Debian development
server\footnote{\url{https://alioth.debian.org}}. About 40 people
signed up to directly contribute to the development, and 15 of these
signed to be responsible for the maintenance of one or more
packages. There is no dedicated Packaging Policy for the Debian Astro
team; mainly the Debian Science Policy \citep{DebianSciencePolicy} is
used as a guide.

\section{Contents}

\begin{figure}[t]
  \includegraphics[height=5cm]{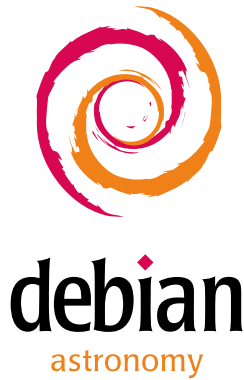}
  \hfill
  \includegraphics[height=5cm]{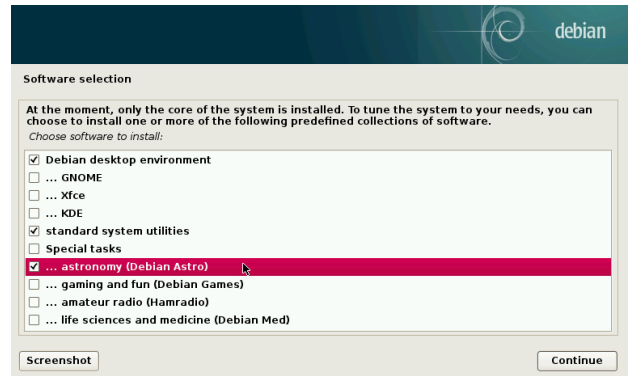}
  \label{fig:installer}
  \caption{{\bf Left:} The logo of the Debian Astro Pure Blend was created by Maria
    Hammerstr{\o}m.
    {\bf Right:} Screenshot of the Debian Stretch installer showing the option to
    select the Debian Astro Pure Blend}
\end{figure}

The packages are organized into 18 sections, where each section is
represented by a Debian meta-package, triggering the installation of the
contained packages as recommendations or suggestions. Selected
packages are:
\begin{itemize}
\item Base libraries: cfitsio, ccfits, qfits, wcslib, wcstools, ESO cpl,
  Starlink AST, PAL, healpix, erfa, giza
\item Common science: fft, cminpack, lapack, numpy, scipy
\item Data Reduction: Astrometry.net, ESO pipelines, Astromatic software
\item Python (2 and 3): Astropy and its affiliated packages, sunpy
\item Virtual Observatory: MOC (Java+Python), Savot, Samp, Aladin, pyvo
\item Radio Astronomy: Casacore, cassbeam, aoflagger, wsclean, purify
\item Viewers: SAOImage DS9, Fv, Ginga, Glue, QFitsView, Skycat, yt
\item Legacy: ESO-MIDAS, Tcl/Tk, GDL Astrolib
\item Education and amateur astronomy: Stellarium, kstars, Indi, Gcx
\end{itemize}

IRAF is not included in Debian for two reasons: First, it contains
code from the Numerical Recipes book, which is not distributable
\citep{Press:1993:NRF:563041}. This effectively makes IRAF non-free
software, despite its licensing claims.  Also the internal structure
of IRAF makes it difficult to follow Debian standards in terms of
compilation, file system hierarchy and portability.

\section{Advantages of packaging for Debian}

Creating official Debian packages has a benefit not only for the users
getting the software in a comfortable way. The major advantage for the
developer is that the software and its integration into the operating
system is extensively tested.

Usually, software packaged for Debian will be built automatically on
all 22 available platforms with different kernels and CPU types, which
is an extensive test of the software for portability. If the software
contains build time tests, they are executed on each platform as
well. Even if the use of non-x86 platforms is currently small, this
ensures reliability to future hardware development.

For each package, continuous integration tests can be setup, which are
then executed within the Debian infrastructure on every dependency
change. Currently, the majority of the packages in the Debian Astro
Pure Blend uses these CI tests, which give an early warning on
incompatible changes of the whole software ecosystem. Also there are
frequent rebuilds of the whole Debian archive from source. Several
Debian specific QA tests are continuously applied to the whole
archive, reporting issues from spelling errors to important compiler
warnings.  The results from all these tests can be propagated back to
the upstream authors and help managing a good quality of their
software.

Due to the highly standardized packaging, Debian is also a popular
system among software engineers doing research in topics like quality
assurance, software metrics, compilers. This leads to very detailed
bug reports, which also can help upstream to improve the software.

Packages in Debian are tightly coupled to the development of the
Debian distribution. Dependencies are automatically recognized. For
example, if a shared library changes the binary interface in a new
version, all dependent packages are automatically recompiled against
the new library. The dependency system also avoids a silent removal of
packages if they are needed by other packages. This is especially
important for astronomy software, since it may depend on very specific
and sometimes ancient code which otherwise could be erased from the
distribution.

Since Debian is the base for a family of Linux distributions, all
packages that are included in Debian are automatically migrated to the
derivative distributions. In this way, Debian Astro and its packages
will be available on Ubuntu, Mint, and other important Linux
distributions.

Uploading packages to Debian has a self-magnifying effect: a strong
Debian Astro Pure Blend will attract other people to contribute their
own packages. Debian follows a ``Bazaar'' development style
\citep{Raymond:1999:CB:580808}. The development is open and
transparent, so everyone can follow and contribute with bug reports,
bug fixes or patches.

When a package cannot be maintained by its original maintainer, it
will still get some attention: First, the Debian Astro team will try
to keep the package in a good shape. The package can then be adopted
by another maintainer. Even when a package is completely orphaned, ir
will get some maintenance: Debian developers can directly upload fixes
if the package has a release critical error. And Debian has a
dedicated quality assurance team that fixes the most urgent problems.

In the past, there were several attempts to build Debian packages for
required software locally. Often this results in the duplication of
work and technical conflicts due to the uncoordinated
development. Building official Debian packages helps to create a
homogeneous software base which is usable for the whole community.

\section{Release}

Since the Debian Astro Pure Blend is integrated into Debian, there are
no separate releases of Debian Astro. We plan to release the first
regular version of the Debian Astro Pure Blend as part of Debian 9.0
(``Stretch''). It is estimated that Debian Stretch will be released in
2017. We plan to allow the selection of Debian Astro already during
the installation of the distribution.

Ubuntu and other distributions deriving from the developer version of
Debian include snapshots of Debian Astro since 2016 (Ubuntu release
16.04 LTS ``Xenial Xerus'').
\section{Acknowledgments}

We thank all contributors to Debian Astro and Debian in general. This
includes the upstream authors as well as the maintainers of the
packages, but also those who reported bugs, submitted patches or
otherwise donated their resources to the project.

\bibliography{P2-20}

\end{document}